# Detection of vascular leukoencephalopathy in CT images

Zuzana Cernekova[0000−0002−7617−4192], Viktor Sisik, and Fatana Jafari[0009−0007−0758−735X]

Faculty of Mathematics Physics and informatics, Comenius University Bratislava, Slovakia {zuzana.cernekova,sisik1,fatana.jafari}@uniba.sk

**Abstract.** Artificial intelligence (AI) has seen a significant surge in popularity, particularly in its application to medicine. This study explores AI's role in diagnosing leukoencephalopathy, a small vessel disease of the brain, and a leading cause of vascular dementia and hemorrhagic strokes. We utilized a dataset of approximately 1200 patients with axial brain CT scans to train convolutional neural networks (CNNs) for binary disease classification. Addressing the challenge of varying scan dimensions due to different patient physiologies, we processed the data to a uniform size and applied three preprocessing methods to improve model accuracy. We compared four neural network architectures: ResNet50, ResNet50 3D, ConvNext, and Densenet. The ConvNext model achieved the high- est accuracy of 98.5% without any preprocessing, outperforming models with 3D convolutions. To gain insights into model decision-making, we implemented Grad-CAM heatmaps, which highlighted the focus areas of the models on the scans. Our results demonstrate that AI, particularly the ConvNext architecture, can significantly enhance diagnostic accu- racy for leukoencephalopathy. This study underscores AI's potential in advancing diagnostic methodologies for brain diseases and highlights the effectiveness of CNNs in medical imaging applications.

## 1 Introduction

The rapid advancement and integration of artificial intelligence (AI) in various fields have notably impacted medicine. AI technologies are transforming diagnostic approaches and patient care, offering new insights into complex medical conditions. One such condition is leukoencephalopathy, a disease of small brain vessels also known as cerebral microangiopathy. This condition is the most common cause of vascular dementia and a major contributor to hemorrhagic strokes, necessitating effective diagnostic tools. Figure 1 illustrates two CT brain scans: one with marked leukoencephalopathy and a comparative slice without it.

Current diagnostic methods for leukoencephalopathy rely heavily on brain imaging techniques such as computed tomography (CT) scans. However, the interpretation of these scans requires significant expertise and can be time-consuming. AI, particularly convolutional neural networks (CNNs), offers a promising solution by automating and potentially improving the accuracy of such diagnoses.

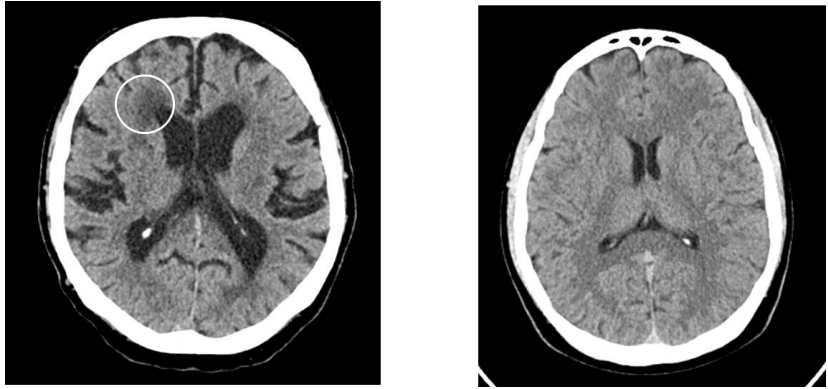

**Fig. 1.** Leukoencephalopathy on a CT scan - the left image shows a brain slice with leukoencephalopathy marked in a white circle; the right image shows a brain slice without leukoencephalopathy

In this study, we leverage a dataset of approximately 1200 patients with axial brain CT scans to train CNN models for the binary classification of leukoencephalopathy. By preprocessing the data to a uniform size, we address challenges such as varying scan dimensions due to different patient physiologies. Furthermore, we employ three different preprocessing methods to enhance model accuracy and compare the performance of four neural network architectures: ResNet50 [8], ResNet50 3D, ConvNext [11], and Densenet [9].

Our findings indicate that the ConvNext architecture achieved the highest classification accuracy of 98.6% without any preprocessing. We also utilized Grad-CAM to generate heatmaps, providing insights into the regions of the scans that the models focused on during classification. This research underscores the potential of AI in advancing diagnostic methodologies for brain diseases, particularly leukoencephalopathy, and sets the stage for further exploration and refinement of AI-driven diagnostic tools in clinical settings.

## 2  Related work

Leukoencephalopathy is an active research area. Recent 2024 studies include neurosurgical perspectives on cerebral calcifications and cysts [16], atypical MRI features in progressive multifocal leukoencephalopathy, and advanced imaging techniques for chemoradiotherapy-induced leukoencephalopathy [4].

## 2.1 Medical Image Processing

In this section, we review previous approaches and studies in the field of artificial intelligence that have influenced our work. The article [6] addresses the classification of brain CT scans into hemorrhagic, ischemic, and normal categories. It tackles two main areas: image preprocessing and image classification using neural networks.

The proposed approach for medical image preprocessing (CT slices) focuses on removing contrast abnormalities to improve classification accuracy. This involves creating two copies of input images, performing contrast adjustments for better visualization on the first copy, and applying average filtering on the second. The preprocessed images are then merged together to form a single image. Regarding model architectures, the article introduces a newly proposed architecture of convolutional neural networks called P_CNN. Unlike other deep learning architectures, P_CNN can process CT scan images without resizing them, which is crucial for preserving image quality. The architecture includes the use of 96 filters in the second layer, convolution with the input layer, ReLU activation, and max pooling. The article also compares P_CNN with other CNN architectures such as AlexNet and ResNet50.

Overall, this approach involves a detailed algorithm for image preprocessing and the use of the P_CNN framework for image classification.

In the article [17], the authors described a model for lung cancer classification trained on CT scans. They outlined the process of working with CT scans in DICOM format and their conversion into Hounsfield units.

Article [15] focuses on developing deep convolutional neural networks for detecting COVID-19 from medical images, proposing a single architecture for both CT scans and X-ray images simultaneously. The authors discuss the importance of rapid and reliable COVID-19 detection, emphasizing the need for effective tools for the diagnosis and monitoring of this disease. They highlight the significance of comparing COVID-19 with other coronavirus diseases to better understand its characteristics and spread. The article describes a proposed deep neural network, experimentation methods, and achieved results, focusing on optimizing parameters for the best model performance.

The proposed neural network in the article consists of 3 alternating convolutional and pooling layers. The convolutional layers had 32, 16, and 8 filters with sizes of $5 \times 5$, $4 \times 4$, and $3 \times 3$, respectively, using ReLU activation.

Within the article, the authors compared their proposed model with other architectures like InceptionV3, MobileNet, and ResNet, where their model outperformed all others with an accuracy of 96.28%. Additionally, their proposed model had significantly fewer parameters compared to other model architectures. In further research [19], a convolutional neural network was proposed for the binary classification of chest CT scans into COVID-19 positive and negative cases. The dataset used in the article consisted of 746 CT scans collected from 216 patients, including 349 images from COVID-19-infected patients and 397 from non-infected patients. The images varied in dimensions, with heights ranging from 153 to 1853 and widths from 153 to 1485. These images were collected by a radiologist in

Wuhan during the COVID-19 outbreak from January to April 2020 and are publicly available for research.

For model training, all images were resized to a uniform size of 150 × 150 pixels and labeled according to classes 0 (negative) and 1 (positive). The entire dataset was then split into a training subset comprising 80% of the dataset and a testing subset comprising 20%. Both divided datasets were normalized between values of 0 − 1.

The authors employed algorithms for binary classification using CNNs with hyperparameters to achieve higher accuracy in detecting COVID-19. The algorithm involved tuning hyperparameters such as different numbers of epochs, batch sizes, and various optimizers. The trained model with the best parameters achieved an accuracy of 86.9%.

## 3 Dataset

We have acquired a dataset of CT scans from the hospital of St. Cyril and Methodius in Bratislava. The dataset consisted of 1244 folders, each named with a numerical identifier (ID) representing anonymized patients. Each patient's folder contained .DCM (DICOM) files for individual scan slices and a .json file with the same numerical identifier, which included metadata such as medical findings, examination number, and the doctor's name. This study focused exclusively on brain slices in a specific plane, filtering out other body parts or planes.

Medical imaging data, particularly CT scans, are usually stored in DICOM (Digital Imaging and Communications in Medicine) format. This standard is widely used in modern medical imaging devices due to its ease of integration and continuous development. DICOM files, represented as ".dcm", consist of a header and image data encapsulated in one file. The header contains patient demographics, acquisition parameters, image dimensions, and intensity data necessary for proper image display. This encapsulation ensures that image data cannot be separated from the header, maintaining the integrity and context of the image.

CT scan data are expressed in Hounsfield Units (HU), which are linear transformations of measured X-ray absorption coefficients relative to water. These units serve as gray levels in the voxels of CT images. A voxel, or volumetric pixel, is a data point on a three-dimensional grid. In CT imaging, voxels represent the varying densities within the scanned volume, providing a detailed three-dimensional representation of the scanned area. Bones appear lighter on CT images due to their higher density and greater radiation absorption, while water and air appear darker. The standard conversion formula for calculating HU for any material is:

$$HU = \frac{\mu - \mu_{water}}{\mu_{water} - \mu_{air}} \times 1000 \qquad (1)$$

where $\mu$ is the absorption coefficient of the examined region.

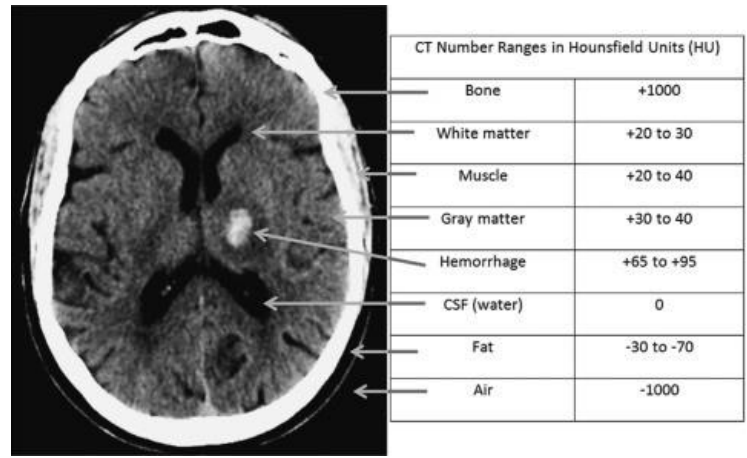

**Fig. 2.** Example of Hounsfield Unit values in different brain regions. From top to bottom, the values are for bone, white matter, muscle, gray matter, hemorrhage, water, fat, and air [10].

## 4 CNN Architectures

Convolutional Neural Networks (CNNs) are the cornerstone of many successful applications in image processing. Among the most well-known architectures are *ResNet50*, *DenseNet*, *VGGNet*, and *ConvNeXt*, all of which have significantly advanced the field of computer vision. Below, we briefly describe the architectures utilized in this work.

**ResNet50** is a foundational CNN architecture that employs residual blocks to enhance the efficiency of deep learning. This architecture includes 50 layers, comprising convolutional and fully connected layers. Residual blocks enable information to be transmitted through multiple layers of the network without loss or learning difficulties. This approach helps address the vanishing gradient problem in deep neural networks, where adding more layers can lead to poorer performance due to learning issues [7].

**DenseNet** is another prominent CNN architecture consisting of transition layers and dense blocks. Each convolutional layer in a dense block is connected to all other layers within the block. This unique mechanism enhances the network's learning capacity by repeatedly leveraging features and reducing parameter requirements, thereby improving gradient flow during training [20].

**ConvNeXt** (Convolutional Neural Network eXtended) represents an innovative architecture that significantly pushes the boundaries of CNNs. ConvNeXt combines CNNs and transformers to leverage the strengths of both architectures. It employs bottleneck layers inspired by transformer architecture, where the number of elements is downsampled and upsampled, along with deep convolutions and residual blocks. Unlike traditional CNNs that heavily use Rectified Linear Units (ReLU), ConvNeXt replaces ReLU with Gaussian Error Linear

Units (GeLU). The hybrid nature of ConvNeXt allows it to achieve superior results in various computer vision tasks [2, 12].

These architectures represent a fundamental part of the rich spectrum of CNNs that have a significant impact on the field of computer vision. Each has unique features and advantages tailored to different tasks and application needs in image processing.

All the mentioned architectures are typically designed with *2D* convolutional layers; however, in our work, we also explored their *3D* convolutional counterparts.

CNN architectures may have specific requirements for minimum image size due to their configuration and filter schemes, which are critical for their effi- ciency and performance. Filtering local parts of an image with small kernels in convolutional layers can be more challenging with smaller dimensions, potentially degrading the network's ability to capture relevant patterns [1].

## 5 Data preprocessing

In this session, we will describe data preprocessing done before training the CNN model. Data preprocessing is essential in preparing medical imaging data for machine learning models. This process ensures that the data is standardized and optimized for effective model training and analysis. In our study, we adopted several preprocessing steps to enhance the quality and consistency of the input data.

### 5.1 Depth normalization

Initially, all brain images of each patient were standardized to a uniform $depth$, which represents the number of stacked 2D slices in the final 3D image. The variation in depth is due to differences in head sizes among patients and variations in the determination of scan start/end points by technicians before each CT scan (initial images typically show only air-black).

The central slice and the surrounding slices were chosen based on the $depth$ and $center\_position$ parameters as follows: We set the $center\_position$ of the new volume to $\frac{2}{5}$ of the total number of slices $n$ from the start of the skull to the neck. Therefore, the central $p$-th slice is calculated as $p = n \times center\_position$. We then select $p \pm \frac{depth}{2}$ slices from this central slice. In our settings, $depth$ was set to 30 slices.

### 5.2 Resizing Scans

Then we resize the images to ensure uniform dimensions, a crucial step for machine learning models requiring fixed-size inputs. In our study, most patient scan slices had dimensions of 512 × 512 × $depth$, with depth varying per patient due to physiological differences. However, we encountered instances where scan dimensions were 512 × **600** × $depth$. These outliers needed to be resized to match the majority dimension to facilitate their inclusion in tensors for further processing.

Beyond ensuring uniform dimensions, resizing images is crucial for computational efficiency. Reducing the size of large images can significantly decrease the computational resources required for training and inference in machine learning models. Smaller images typically lead to faster training times and lower memory consumption, which is especially beneficial when working with large datasets. Therefore, we reduced the image size to 256 × 256, optimizing for consistency and computational efficiency.

### 5.3 Pixel value Rescaling

We selected only the pixel values ($x$) from the interval $< 0, 100 >$ and normalized them to the range $< 0, 1 >$ using the formula

$$x = \begin{cases} 1 & if\ x \geq 100 \\ 0 & if\ x < 0 \\ x/100 & else \end{cases}$$

(2)

### 5.4 Preprocessing A

In our study, we utilized different combinations of preprocessing methods. In the first approach, we applied only the common preprocessing steps mentioned above.

### 5.5 Preprocessing B - with Filtering and Morphological Operations

In the second approach, we added the following image processing steps in addition to the common preprocessing steps:
- Filtering: We removed pixels that represented calcifications, skull, and water, following the formula:

$$x = \begin{cases} 0 & if\ x \geq 0.8 \\ 0 & if\ x \leq 0.18 \\ 0 & oterwise \end{cases}$$

(3)

- Morphological operation: We applied the morphological opening operation with a structuring element of size 4 × 4. This operation helps reduce noise by performing erosion followed by dilation, thus enhancing the target areas.

### 5.6 Preprocessing C - with Mean Filtering and Contrast Adjustment

In the third approach, we applied mean filtering and contrast adjustment in addition to the common preprocessing steps.

- Mean filtering: We applied a mean filter with a kernel size of 3 × 3 to reduce noise and smooth the images.

- Contrast adjustment: We adjusted the image contrast by stretching the intensity range between the darker regions (pixel values around $low$ = 0.15) and the lighter regions (pixel values around $high$ = 0.65). This technique, inspired by *skimage's* rescale intensity function, enhances the visibility of critical structures within the scans.
- Filtering: We removed all pixels with a value of $x$ = 1 to eliminate irrelevant areas such as the skull.

### 5.7 Data Augmentation

Data augmentation is a crucial technique in training convolutional neural networks (CNNs), which are known for their high data requirements. It involves applying diverse transformations to existing images to enhance the diversity of the training dataset. This helps CNNs generalize better to different variations in input data and improves their performance on unseen test data. Augmentation is especially beneficial when the training dataset is limited in size, effectively increasing its size and reducing overfitting. Common augmentations include rotation, scaling, flipping, and adding noise [5, 13].

In our study, we applied rotation to simulate variations in patient positioning during scans, ensuring our models learn robust features. We also used horizontal flipping to account for the absence of statistically dominant findings of leukoencephalopathy on a specific side of the brain. These augmentations aimed to enhance the model's ability to generalize across different orientations and conditions encountered in medical imaging analysis.

## 6 Training Details and Obtained Results

We divided the data into three subsets without further modifications: training, validation, and testing in the ratios 70 : 15 : 15. Class balance was maintained naturally as the dataset contained a near 1:1 ratio of patients with leukoencephalopathy (633) and without (611). No additional class weighting was applied during training, but we ensured an equal distribution across the training, validation, and test sets.

The architectures we compared were ResNet50, ResNet50 3D, ConvNext DenseNet. For training **2D models**, we identified the following optimal hyperparameters: a learning rate of $10^{-5}$, the Adam optimizer, a batch size of 64, and 50 epochs.

For training the **3D model**, the optimal hyperparameters were a learning rate of $10^{-5}$, the Adam optimizer, a batch size of 2, and 50 epochs.

During each training session, we utilized *early stopping* when the model's performance ceased to improve after several epochs.

For our training, we used pre-trained models with weights from the *ImageNet* dataset, which consists of color images with 3 channels. To match this format, we replicated our grayscale data three times to create the necessary number of channels.

A significant issue encountered was *overfitting* during the training of 3D models. Approximately 1000 data samples were used for training, and the model struggled to generalize well on new data.

## 6.1 Evaluation

In this section, we evaluate the performance of model architectures on the pre-processed data. We used binary accuracy as our accuracy metric, which measures how often the predicted values $y_{pred}$ match the actual values $y_{true}$. Mathematically, it is defined as:

$$binary_{accuracy} = \frac{count(y_{pred} == y_{true})}{n} \qquad (4)$$

where $n$ is the total number of elements in the test set. We used binary cross-entropy (BCE) as the loss function, which is commonly used in binary classification problems. BCE measures the dissimilarity between the true classes and the predicted ones.

It is important to note that we evaluated the classification accuracy of the models on individual slices in the case of 2D models and on selected slices for the 3D models. The slice selection method is described in Section 5.1. This means we did not evaluate the accuracy of classifying individual patients based on their CT scans.

It is important to clarify that our evaluation focused on the classification accuracy of individual CT slices rather than entire patient scans. For 2D models, we assessed accuracy on each slice separately, while for 3D models, only a selected subset of slices was evaluated, as described in Section 5.1. This approach means that patients could be represented by multiple slices, potentially leading to multiple evaluations for a single patient. Thus, the reported accuracy reflects slice-level performance, not patient-level diagnosis.

The results of training various model architectures using data preprocessing type A, B, and C are shown in Table 1.

**Table 1.** Results on validation data set preprocessed using methods A, B and C

|  | Model | ResNet50 | ResNet50 3D | ConvNeXt | DenseNet |
|---|---|---|---|---|---|
| preprocessing A | AVG | 92.5% | 75.7% | 95.9% | 93.3% |
|  | BEST | 95.7% | 79.0% | **98.6%** | 94.5% |
| preprocessing B | AVG | 85.5% | 74.4% | 87.3% | 84.0% |
|  | BEST | 86.0% | 79.0% | 88.2% | 85.9% |
| preprocessing C | AVG | 90.9% | 74.4% | 86.4% | 89.4% |
|  | BEST | 92.5% | 76.6% | 89.6% | 90.3% |

The recorded results were obtained from the classification accuracy on the validation set, which matched the accuracy on the test data within a deviation of ±3%. The evaluation of the models consisted of more than 8 trained models in each case, meaning that the AVG row in the tables represents the average accuracy results of the models on the validation data from more than 8 different training sessions of the same model and training parameters. The BEST row indicates the best results among these training sessions.

Table 2. Results on test data set preprocessed using methods A, B and C

|  | Model | ResNet50 | ResNet50 3D | ConvNeXt |
|---|---|---|---|---|
| preprocessing A | AVG | 94.1% | 71.7% | 92.6% |
| preprocessing A | BEST | 94,7% | 77.6% | **98.5%** |
| preprocessing B | AVG | 88.8% | 75.2% | 87.3% |
| preprocessing B | BEST | 91.4% | 75.8% | 88.1% |
| preprocessing C | AVG | 87.2% | 74.8% | 87.2% |
| preprocessing C | BEST | 88.2% | 78.3% | 89.3% |

The results on the test set are shown in Table 2. Models of all architectures achieved the best results without any data preprocessing. The overall best-performing model, with an accuracy of 98.5% on the test set, was the ConvNeXt architecture.

A few words on model accuracy. The trained models aim not to replace doctors but to provide an objective tool for aiding diagnosis. Therefore, the model should ideally have as few false negatives as possible, even at the cost of higher false positives, which a doctor can then review. The problem arises when the model fails to identify a patient who should be classified as positive.

**6.2 Discussion**

The main issue in this study was the small dataset specific to the problem. Training models with architectures using 3D convolutional layers often led to overfitting, even with increased regularization or data augmentation.

Another significant challenge was the selection of slices (images) for training the neural network. Only a few slices from the entire scan contain areas of leukoencephalopathy. Finding a general rule for selecting these slices is difficult due to different brain physiologies and associated conditions (e.g., brain atrophy). Strict rules often eliminate useful slices, resulting in a very small trained set. Looser rules included slices without leukoencephalopathy, potentially skewing model accuracy.

In 2D models, a voting system could be introduced to determine the final classification of a patient, mitigating the issue of "poorly" selected slices mentioned above. The voting would involve dividing the entire CT scan into $k$ 2D slices, classifying each, and setting a threshold $t \in <1, k>$. If the number of positively classified slices $p$ exceeds $t$, the patient would be classified as positive.

## 7 Grad-CAM

Grad-CAM, short for Gradient-weighted Class Activation Mapping, is a technique used to visualize which parts of an image are most important for a convolutional neural network (CNN) to make predictions about the class. Grad-CAM operates by computing gradients of the score (the network assigns a score to each class - a probability) concerning the last convolutional layer. These gra- dients are then used to generate a *heatmap* that highlights the regions of the image contributing the most to the prediction of a specific class. In other words, it projects a heatmap onto the image indicating where the model is "looking" when making its prediction.

Grad-CAM is used to enhance the interpretability of CNN models by helping understand the decisions made by the model during prediction through the visualization of regions of interest.

In medical practice, the advantages of Grad-CAM can provide an objective tool to obtain a second opinion on medical data. It assists doctors in highlighting important areas, thereby increasing diagnostic accuracy. Moreover, it supports medical education by illustrating how CNNs analyze images, aiding students in understanding diagnostic thinking [3, 14, 18].

In the case of leukoencephalopathy, the model must focus primarily on the areas of the brain's ventricles, where the problem is located, rather than on the periphery of the brain. This is because an estimated 80-90% of patients exhibit associated brain atrophy alongside leukoencephalopathy. By utilizing Grad- CAM, it is possible to evaluate the model more comprehensively from a different perspective beyond just classification accuracy on test data. In Figures 3 and 4 we can observe that the networks focus on the regions where leukoencephalopathy typically appears.

## 8 Conclusion and Future Work

This study aimed to explore methods for processing medical data, specifically CT scans, and to train an AI model capable of accurately classifying CT slices containing leukoencephalopathy. Developing well-annotated datasets is crucial for training models that can effectively integrate AI into daily medical practice, potentially saving time for healthcare professionals and improving patient outcomes.

We tested three different preprocessing approaches for CT scans, finding that the approach with minimal preprocessing was the most effective. We employed four different model architectures: ResNet50, DenseNet, ConvNeXt, and ResNet50 with 3D convolutional layers. ConvNeXt achieved the highest accuracy of 98.5% on the test set. Although several trained models demonstrated satisfactory accuracy, the limited dataset size was a constraint, particularly for models with 3D convolutional layers, which frequently encountered overfitting issues. The ResNet50 variant with 3D layers showed lower accuracy, but it holds the potential for better capturing the deeper connections between CT scan slices with further data and refinement.

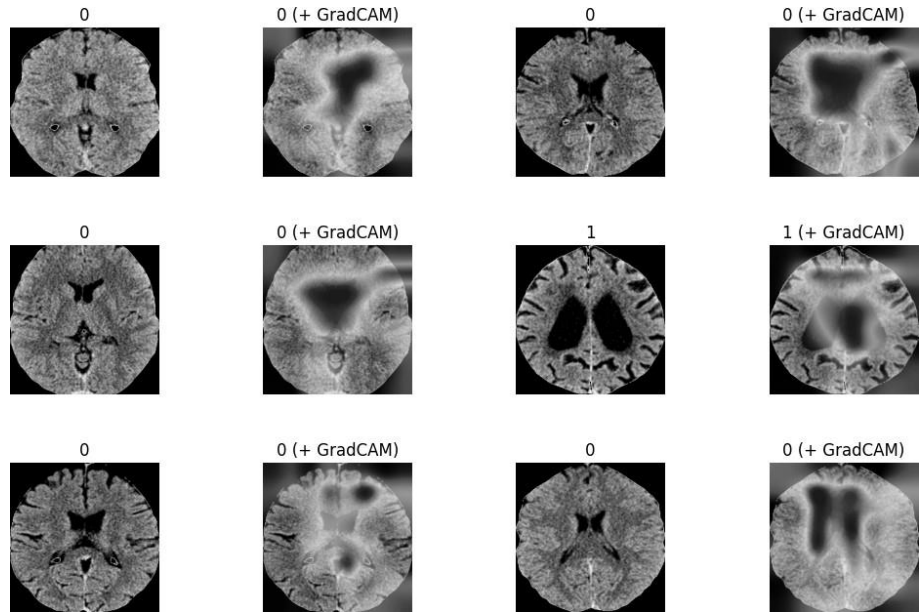

**Fig. 3.** Grad-CAM on test data for a ConvNeXt model trained on preprocessed Type C data is visualized in four columns. The first column displays the test data with their respective classes in the header (0 - negative, 1 - positive). The second column overlays the image with a heatmap (darker areas indicate higher attention), highlighting the regions the model focuses on. The header of this column states the predicted class of the image by the trained network. The same format applies to the third and fourth columns.

In our dataset, an estimated 80-90% of patients had comorbid brain atrophy, visible on CT scans as dark protrusions around the brain's perimeter. We used Grad-CAM heatmaps to highlight the regions where the model focused during classification, providing an additional layer of verification for the model's accuracy.

Future research should focus on expanding the dataset to improve the robustness and generalizability of the models, particularly those utilizing 3D convolutional layers. Investigating advanced preprocessing techniques and their impact on model performance will be essential. Additionally, developing methods for better slice selection and classification, including the use of ensemble techniques and voting mechanisms, could enhance model accuracy and reliability.

**Acknowledgments.** This publication was supported by projects: the TERAIS project, Horizon-Wider-2021 programme of the European Union under grant agreement number 101079338, project APVV-23-0250, and student project UK/3190/2024.

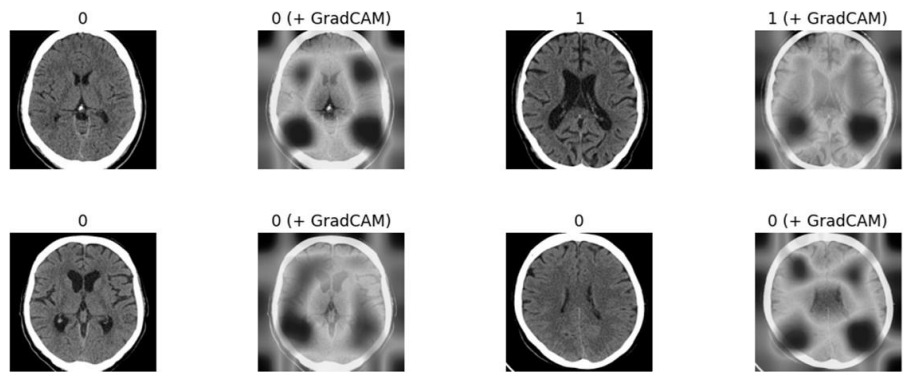

**Fig. 4.** Grad-CAM on test data for a ResNet50 model trained on preprocessed Type A data is visualized in four columns. The first column displays the test data with their respective classes in the header (0 - negative, 1 - positive). The second column overlays the image with a heatmap (darker areas indicate higher attention), highlighting the regions the model focuses on. The header of this column states the predicted class of the image by the trained network. The same format applies to the third and fourth columns.

**Disclosure of Interests.** The authors have no competing interests to declare that are relevant to the content of this article.

## References

1. Adaloglou, N.: Understanding the receptive field of deep convolutional networks. https://theaisummer.com/ (2020), https://theaisummer.com/receptive-field/
2. Ahmadian, A., Liu, L.S., Fei, Y., Plataniotis, K.N., Hosseini, M.S.: Pseudo-inverted bottleneck convolution for darts search space. In: 2023 IEEE International Con- ference on Acoustics, Speech and Signal Processing (ICASSP23). pp. 1–5. IEEE (2023)
3. Aravinda, C., Lin, M., Udaya Kumar Reddy, K., Amar Prabhu, G.: 23 - a demystifying convolutional neural networks using grad-cam for prediction of coron- avirus disease (covid-19) on x-ray images. In: Kose, U., Gupta, D., de Albuquerque, V.H.C., Khanna, A. (eds.) Data Science for COVID-19, pp. 429–450. Academic Press (2021). https://doi.org/https://doi.org/10.1016/B978-0-12-824536-1.00037-X
4. Celardo, G., Scaffei, E., Buchignani, B., Donatelli, G., Costagli, M., Cristofani, P., Canapicchi, R., Pasquariello, R., Tosetti, M., Battini, R., Biagi, L.: Case report: Exploring chemoradiotherapy-induced leukoencephalopathy with 7t imaging and
quantitative susceptibility mapping. Frontiers in Neurology **15**, 1362704 (2024). https://doi.org/10.3389/fneur.2024.1362704
5. Cirillo, M.D., Abramian, D., Eklund, A.: What is the best data augmentation for 3d brain tumor segmentation? In: 2021 IEEE International Conference on Image Processing (ICIP). pp. 36–40 (2021). https://doi.org/10.1109/ICIP42928.2021.9506328

6. Gautam, A., Raman, B.: Towards effective classification of brain hemorrhagic and ischemic stroke using cnn. Biomedical Signal Processing and Control **63**, 102178 (2021)
7. He, K., Zhang, X., Ren, S., Sun, J.: Deep residual learning for image recognition (2015), https://arxiv.org/abs/1512.03385
8. He, K., Zhang, X., Ren, S., Sun, J.: Deep residual learning for image recognition. In: 2016 IEEE Conference on Computer Vision and Pattern Recognition (CVPR). pp. 770–778 (2016). https://doi.org/10.1109/CVPR.2016.90
9. Huang, G., Liu, Z., Van Der Maaten, L., Weinberger, K.Q.: Densely connected convolutional networks. In: 2017 IEEE Conference on Computer Vision and Pattern Recognition (CVPR). pp. 2261–2269 (2017). https://doi.org/10.1109/CVPR.2017.243
10. Kamalian, S., Lev, M.H., Gupta, R.: Chapter 1 - Computed tomography imag- ing and angiography – principles, Handbook of Clinical Neurology, vol. 135. Else- vier (2016). https://doi.org/https://doi.org/10.1016/B978-0-444-53485-9.00001-5, https://www.sciencedirect.com/science/article/pii/B9780444534859000015
11. Liu, Z., Mao, H., Wu, C.Y., Feichtenhofer, C., Darrell, T., Xie, S.: A convnet for the 2020s. In: 2022 IEEE/CVF Conference on Computer Vision and Pattern Recognition (CVPR). pp. 11966–11976 (2022). https://doi.org/10.1109/CVPR52688.2022.01167
12. Liu, Z., Mao, H., Wu, C.Y., Feichtenhofer, C., Darrell, T., Xie, S.: A convnet for the 2020s. In: Proceedings of the IEEE/CVF conference on computer vision and pattern recognition. pp. 11976–11986 (2022)
13. Michelucci, U.: Advanced Applied Deep Learning: Convolutional Neural Networks and Object Detection. Apress (2019), https://books.google.sk/books?id=hJyyDwAAQBAJ
14. Moujahid, Hicham, e.a.: Combining cnn and grad-cam for covid-19 disease predic- tion and visual explanation. Intelligent Automation and Soft Computing **32**(2), 723–745 (2022). https://doi.org/10.32604/iasc.2022.022179
15. Mukherjee, H., Ghosh, S., Dhar, A., Obaidullah, S.M., Santosh, K., Roy, K.: Deep neural network to detect covid-19: one architecture for both ct scans and chest x-rays. Applied Intelligence **51**, 2777–2789 (2021)
16. Novegno, F., Iaquinandi, A., Ruggiero, F., Salvati, M.: Leukoencephalopathy with cerebral calcifications and cysts: The neurosurgical perspective. literature review. World Neurosurgery **190**, 99–112 (2024)
17. Polat, H., Danaei Mehr, H.: Classification of pulmonary ct images by using hybrid 3d-deep convolutional neural network architecture. Applied Sciences **9**(5), 940 (2019)
18. Selvaraju, R.R., Cogswell, M., Das, A., Vedantam, R., Parikh, D., Batra, D.: Grad-cam: Visual explanations from deep networks via gradient-based local- ization. International Journal of Computer Vision **128**(2), 336–359 (10 2019). https://doi.org/10.1007/s11263-019-01228-7
19. Shambhu, S., Koundal, D., Das, P., Sharma, C.: Binary classification of covid-19 ct images using cnn: Covid diagnosis using ct. International Journal of E-Health and Medical Communications (IJEHMC) **13**(2), 1–13 (2021)
20. Yilmaz, F., Kose, O., Demir, A.: Comparison of two different deep learning archi- tectures on breast cancer. In: 2019 Medical Technologies Congress (TIPTEKNO). pp. 1–4 (2019). https://doi.org/10.1109/TIPTEKNO47231.2019.8972042